# The straw man of quantum physics

Peter Morgan
*Physics Department, Yale University, New Haven, CT 06520, USA.*

**One-sentence summary:** The violation of Bell inequalities rules out the straw man class of classical particle property models but leaves untouched the more significant class of classical random field models.

**Abstract:** The violation of Bell inequalities by experiment has convinced physicists that we cannot maintain a classical view of the world. When we argue against the possibility of local realist hidden-variable models, however, the ubiquitous requirement of realism, that "measurement results depend on pre-existing properties of objects that are independent of the measurement"[x], reduces classical theory to a straw man. When our most successful physical theories have been field theories for well over a century, and probabilistic for almost as long, the proper comparison is between quantum fields and random fields, for which there are no sharply defined objects and no properties, so that realism is inapplicable. If we model quantum fluctuations explicitly, we can construct random field models as alternatives to quantum field models.

Bell's first construction of an inequality that must be satisfied by a local realistic theory[1] was refined to apply to classical field theories by considering probability densities for events which occur in spatially separated regions of space-time[2], with a few tricky issues tidied up later[3]. Bell emphasizes his belief that this argument applies to classical fields, not just to particle property models, in Ref. 4. Essentially[5], in order for the observations of a classical field that we make *now* to be rather unusual — in the sense that local realistic hidden-variable models cannot account for them — the classical field would also have to be rather unusual *in the past*. It is usually said, pejoratively, that a "conspiracy" is required, however classical local field models are possible, it is just that the past has to be unusual for the present to be unusual.

A straightforward classical field theory is not mathematically adequate for modern physics because it does not include probabilities, which are needed to model the statistics of experimental results. Bell introduces, in Ref. 2, what he calls a theory of "beables", which he supposes to be a classical statistical theory; however, a well-defined elementary mathematical object, a *random field*, is adequate for his mathematical needs and does not have any awkward ontological connotations. A random field is an indexed set of random variables, the elementary building blocks of mathematical probability theory. The index set can be as simple as {1,2}, representing two dice, say, but the physically interesting case is an index set of functions on space-time, which gives us a *continuous random field*. A quantum field can be presented as a set of measurement operators indexed by a set of functions on space-time; following that lead, a continuous random field can be presented as a commutative quantum field.

There is a substantial logical difference between a local random field model and a local realistic hidden-variable model. For a local random field model, each measurement event is determined by the physical state of the whole experimental apparatus, whereas for a local realistic hidden-variable model measurement results are determined by a point source from which two objects are transmitted. There are no sharply defined pre-existing objects and properties in a random field model to cause measurement events and correlations between them, which are instead caused by correlations of the random field and the detailed statistical mechanics of the experimental apparatus and the actions and reactions between them. A continuous random field model is *contextual* just because the whole experimental apparatus must be modeled.

If the unusual past that is required is thought unreasonable, then quantum field theory has the same problem. Suppose we construct a quantum field state that models the whole of an experimental apparatus that violates Bell inequalities and operators that model statistics of the experimental results. For a sophisticated experiment such as that of Weihs *et al.*[6], which satisfies most physicists that local realistic hidden-variable models are essentially untenable[7], part of the experimental results are the settings of the polarizers that determine what measurements are made; in a detailed model, the measurement settings *are* measurement results, manifestly so in the case of Weihs *et al.*'s experiment, and should be modeled as well as measurements that are more obviously called "results". It may seem excessive to require of a quantum model that the measurement apparatus must be modeled as well as the system that is measured, when a simple appeal to symmetries of the experiment and some standard quantum mechanical calculations yield quite good agreement with experiment, but a detailed model of the thermodynamics of detectors is a necessary aim of quantum field theory if it is to be taken seriously as a more-or-less universal theory. Such a requirement is expressed, for example, by Feynman and Hibbs,

> "The usual separation of observer and observed which is now needed in analyzing measurements in quantum mechanics should not really be necessary, or at least should be even more thoroughly analyzed. What seems to be needed is the statistical mechanics of amplifying apparatus"[8].

Once we construct a quantum field state for a complete experiment, it models not only statistics of the measurement results we observe, but also what statistics of other measurement results we would observe if we were to make different measurements, at earlier times, which would always prefigure the actual measurement statistics we obtain. Hence, unusual statistics of measurement results now no less entail unusual statistics of measurement results in the past and in the future for a quantum field model than they do for a classical random field model.

One concern that Bell and some subsequent authors have emphasized is that the experimenter should be able to make free choices of what measurements will be made (although many physicists react strongly against the idea that an effective physical model should be *required* to include the experimenter). Classical deterministic models for biological systems might prevent the free will of the experimenter, but a model of statistical measures of the choices an experimenter makes does not impinge on their free will for each individual choice. In any case, a comprehensive quantum field model for a Bell inequality-violating experiment models the statistics of measurement settings — the polarizer settings in Weihs *et al.*'s experiment — no more or less than a random field model (which, recall, can be presented as a commutative quantum field), so again this is a problem for a quantum field model as much as for a random field model.

An important feature of random field models for experiments in which quantum effects are significant is the natural appearance of quantum fluctuations. In the random field context quantum fluctuations are differentiated in a fundamental way, by Lorentz invariance, from thermal fluctuations[9]. The amplitude of quantum fluctuations is determined by Planck's constant, just as the amplitude of thermal fluctuations is determined by temperature. A fundamental consequence of the explicit presence of quantum fluctuations in a random field formalism is the concept of a thermodynamic dual of the amplitude of quantum fluctuations, analogously to the concept of entropy as a thermodynamic dual of the amplitude of thermal fluctuations; the interplay of quantum and thermal entropy is accounted for implicitly in existing physical

**Random fields.** We cannot in general measure a quantum field or a continuous random field *at a point* unless we are willing to talk in mathematically undefined terms — when we could say that the field at a point is almost always ±∞ — but we *can* measure the field averaged over a small region of space-time, with a weight function *f*, to give a set of operators $\Phi_f$ indexed by weight functions (which are called *test functions*). These weighted averages of field values that can loosely be thought of as ±∞ at a point are finite [more properly, Φ is a linear functional of test functions *f* to operators $\Phi_f$; there is no infinite field at a point in the theory, there are only finite observations of $\Phi_f$]. The essential difference between a quantum field and a continuous random field, from which all other differences follow, is that $\Phi_f$ and $\Phi_g$, for different test functions *f* and *g*, may be incompatible observables for a quantum field but are always compatible for a continuous random field. This difference of incompatibility or compatibility of measurements is significant and is the basis of a long-running critique of the idea that the violation of Bell inequalities is about locality[10,11,12].

arguments in the quantum field context, whereas the interplay must be explicit in a random field formalism. Random fields have a general requirement to model the effects of quantum fluctuations explicitly, which can be alternately useful or a hindrance. No-one will stop using quantum fields and quantum mechanics whenever that is easier to do, but we can understand quantum theory by understanding the relationship between quantum fields and classical random fields, which is easier than understanding the relationship between quantum mechanics and classical particle property models.

We should think of Feynman and Hibbs' "amplifying apparatus" as *sensitive apparatus*, not as detection or measurement devices or apparatus: it is by now well-established that we are best not to think in terms of systems, objects, or particles that exist independently and can be "detected" or "measured". A "sensitive apparatus" is a thermodynamic system that is delicately engineered in a rest, metastable state, ready to make a transition to a different, precipitated state that can be observed macroscopically. Even when the sensitive apparatus is completely isolated, there will be a "dark-rate" of thermodynamic transitions; when the sensitive apparatus is placed near other experimental apparatus, the average rate and other statistics of the transitions will change, differently for different kinds and geometries of the whole apparatus.

There are many different kinds of sensitive apparatus. A photographic plate, for example, has no feedback loop to return the precipitated state to the ready state, which would make statistics awkward if we could not construct a large-scale structure of many small, effectively isolated thermodynamic systems that are all in the ready state. Semiconductor detectors are carefully engineered to be very small, numerous, with "dead-times" (when the sensitive apparatus is not in the ready state) that are as small as possible, and with sensitivities tuned, as far as possible, to electromagnetic, electronic, muonic, or other fields. We say that the field causes the events, but the field and the sensitive apparatus are both required.

Despite the constant fine-grained transitions of the sensitive apparatus, many classic experiments are engineered to be in a coarse-grained equilibrium state, in the sense that statistics of observed events do not vary over time. For an experimental apparatus that *is* in a coarse-grained equilibrium of the field, the field and the statistics of the observed events are globally determined by the whole experimental apparatus. Changing any part of the apparatus — and waiting for equilibrium to be re-established — will change the observed statistics of events. This classical understanding agrees naturally and firmly with the insistence of the Copenhagen interpretation of quantum mechanics that the whole apparatus affects the results that will be obtained, an idea that is not natural to particle-oriented approaches. This understanding also agrees with Bell's polemic, "Against 'measurement'"[13], which emphasizes the role of the whole experiment and the difficulties of talking about systems, objects, or particles and measurement of their properties.

The stumbling blocks for classical particle property models for Bell inequality-violating experiments have typically been their mathematically ad-hoc nature, nonlocality, contextuality, or the lack of a Lorentz invariant formalism, which certainly make them poor in comparison with quantum mechanics and quantum field theory. Quantum field theory, however, is notorious for its need for renormalization, which is a mathematically dubious enough procedure that no rigorous interacting quantum field models are known in Minkowski space. It has been shown recently, however, that a large class of rigorous interacting continuous random fields can be constructed in a Lorentz invariant formalism[14] — because a commutative algebraic structure is less confining than algebraic constraints on quantum fields — making the mathematics of quantum fields, in this respect, poor in comparison to the mathematics of random fields.

Renormalization is a first reason to justify the use of random fields. Secondly, although accommodations to the measurement problem are by now natural enough that they are of little concern in our daily use of quantum mechanics, the construction of random field models would also lessen interpretational issues. Thirdly, quantum gravity has to some extent stumbled on the conceptual mismatch between quantum field theory and classical general relativity[15], a mismatch that is reduced if we adopt the use of random fields, although the probabilistic aspect that is introduced to classical models by random fields may well cause as many technical difficulties as we find in quantum gravity.

There are several existing classes of models that to varying degrees may be understood to be random fields, including those of Adler[16], 't Hooft[17], Khrennikov[18], Morgan[14], and Wolfram[19], all of which have been relatively little pursued partly because they have been taken to be obviously in conflict with the violation of Bell inequalities. The mathematics of random fields is certainly very little developed as yet, but it is nonetheless time to put away the straw man of classical particle property models.

email: peter.w.morgan@yale.edu
homepage: http://pantheon.yale.edu/~PWM22